\documentstyle[psfig]{mn}

\newif\ifAMStwofonts
\def\lapp{\ifmmode\stackrel{<}{_{\sim}}\else$\stackrel{<}{_{\sim}}$\fi}
\def\gapp{\ifmmode\stackrel{>}{_{\sim}}\else$\stackrel{>}{_{\sim}}$\fi}
\def\rotm{rad~m$^{-2}$}
\def\deg{\ifmmode^{\circ}\else$^{\circ}$\fi}
\def\psr{PSR~B1259$-$63}
\def\psrbe{PSR~B1259$-$63/SS2883}
\def\lsi{LSI+61\degr303}
\def\be{SS2883}
\def\pria{B0823$-$500}
\def\prib{B1934$-$638}
\def\seccal{B1251$-$713}
\def\beq{\begin{equation}}
\def\eeq{\end{equation}}
\def\peri{\ifmmode{\cal{T}}\else${\cal{T}}$\fi}
\newcommand{\p}[1]{\ifmmode{\peri #1}\else$\peri #1$\fi}
\def\rdot{\ifmmode{R_\odot}\else$R_\odot$\fi}
\def\mdot{\ifmmode{M_\odot}\else$M_\odot$\fi}
\def\ldot{\ifmmode{L_\odot}\else$L_\odot$\fi}
\def\rstar{\ifmmode{R_*}\else$R_*$\fi}
\def\mstar{\ifmmode{M_*}\else$M_*$\fi}
\def\lstar{\ifmmode{L_*}\else$L_*$\fi}
\newcommand{\pow}[2]{\ifmmode{{#1}^{#2}}\else{#1}$^{#2}$\fi} 
\newcommand{\prog}[1]{\textsc{#1}}
\def\mir{\prog{Miriad}}
\ifoldfss
  \ifCUPmtlplainloaded \else
    \NewTextAlphabet{textbfit} {cmbxti10} {}
    \NewTextAlphabet{textbfss} {cmssbx10} {}
    \NewMathAlphabet{mathbfit} {cmbxti10} {} % for math mode
    \NewMathAlphabet{mathbfss} {cmssbx10} {} %  "   "    "
  \fi
  \ifAMStwofonts
    \ifCUPmtlplainloaded \else
      \NewSymbolFont{upmath} {eurm10}
      \NewSymbolFont{AMSa} {msam10}
      \NewMathSymbol{\upi}     {0}{upmath}{19}
      \NewMathSymbol{\umu}     {0}{upmath}{16}
      \NewMathSymbol{\upartial}{0}{upmath}{40}
      \NewMathSymbol{\leqslant}{3}{AMSa}{36}
      \NewMathSymbol{\geqslant}{3}{AMSa}{3E}

    \fi
  \fi
\fi % End of OFSS

\ifnfssone
  \newmathalphabet{\mathit}
  \addtoversion{normal}{\mathit}{cmr}{m}{it}
  \addtoversion{bold}{\mathit}{cmr}{bx}{it}
  \newmathalphabet{\mathbfit} % math mode version of \textbfit{..}
  \addtoversion{normal}{\mathbfit}{cmr}{bx}{it}
  \addtoversion{bold}{\mathbfit}{cmr}{bx}{it}
  \newmathalphabet{\mathbfss} % math mode version of \textbfss{..}
  \addtoversion{normal}{\mathbfss}{cmss}{bx}{n}
  \addtoversion{bold}{\mathbfss}{cmss}{bx}{n}
  \ifAMStwofonts
    \ifCUPmtlplainloaded \else
      \UseAMStwoboldmath
      \makeatletter
      \new@mathgroup\upmath@group
      \define@mathgroup\mv@normal\upmath@group{eur}{m}{n}
      \define@mathgroup\mv@bold\upmath@group{eur}{b}{n}
      \edef\UPM{\hexnumber\upmath@group}
      \new@mathgroup\amsa@group
      \define@mathgroup\mv@normal\amsa@group{msa}{m}{n}
      \define@mathgroup\mv@bold\amsa@group{msa}{m}{n}
      \edef\AMSa{\hexnumber\amsa@group}
      \makeatother
      \mathchardef\upi="0\UPM19
      \mathchardef\umu="0\UPM16
      \mathchardef\upartial="0\UPM40
      \mathchardef\leqslant="3\AMSa36
      \mathchardef\geqslant="3\AMSa3E
    \fi
  \fi
\fi % End of NFSS release 1

\ifnfsstwo
  \DeclareMathAlphabet{\mathbfit}{OT1}{cmr}{bx}{it}
  \SetMathAlphabet\mathbfit{bold}{OT1}{cmr}{bx}{it}
  \DeclareMathAlphabet{\mathbfss}{OT1}{cmss}{bx}{n}
  \SetMathAlphabet\mathbfss{bold}{OT1}{cmss}{bx}{n}
  \ifAMStwofonts
    \ifCUPmtlplainloaded \else
      \DeclareSymbolFont{UPM}{U}{eur}{m}{n}
      \SetSymbolFont{UPM}{bold}{U}{eur}{b}{n}
      \DeclareSymbolFont{AMSa}{U}{msa}{m}{n}
      \DeclareMathSymbol{\upi}{0}{UPM}{"19}
      \DeclareMathSymbol{\umu}{0}{UPM}{"16}
      \DeclareMathSymbol{\upartial}{0}{UPM}{"40}
      \DeclareMathSymbol{\leqslant}{3}{AMSa}{"36}
      \DeclareMathSymbol{\geqslant}{3}{AMSa}{"3E}
    \fi
  \fi
\fi % End of NFSS release 2

\ifCUPmtlplainloaded \else
  \ifAMStwofonts \else % If no AMS fonts
    \def\upi{\pi}
    \def\umu{\mu}
    \def\upartial{\partial}
  \fi
\fi

\title{The 2000 Periastron Passage of PSR B1259--63}
\author[Connors, Johnston, Manchester, McConnell]
{T.~W.~Connors$^{1}$\thanks{Present address: Swinburne Centre for 
Astrophysics and Supercomputing, Swinburne University of Technology,
Hawthorn, Vic 3122, Australia},
S.~Johnston$^{1}$, R. N. Manchester$^{2}$ and D. McConnell$^{2}$\\
$^1$School of Physics, University of Sydney, NSW 2006, Australia\\
$^2$Australia Telescope National Facility, CSIRO, PO Box 1710, Epping NSW
2121, Australia
}

\date{\today}

\pagerange{\pageref{firstpage}--\pageref{lastpage}}
\pubyear{2002}

\begin{document}

\maketitle

\label{firstpage}

\begin{abstract}
  We report here on a sequence of 28 observations of the binary pulsar
  system \psrbe{} at four radio frequencies made with the Australia
  Telescope Compact Array around the time of the 2000 periastron
  passage.  Observations made on 2000 Sep 1 show that the pulsar's
  apparent rotation measure (RM) reached a maximum of $-14800 \pm
  1800$~\rotm{}, some 700 times the value measured away from
  periastron, and is the largest astrophysical RM measured. This
  value, combined with the dispersion measure implies a magnetic field
  in the Be star's wind of 6~mG.  We find that the light curve of the
  unpulsed emission is similar to that obtained during the 1997
  periastron but that differences in detail imply that the emission
  disc of the Be star is thicker and/or of higher density. The
  behaviour of the light curve at late times is best modelled by the
  adiabatic expansion of a synchrotron bubble formed in the
  pulsar/disc interaction.  The expansion rate of the bubble
  $\sim$12~km~s$^{-1}$ is surprisingly low but the derived magnetic
  field of 1.6~G close to that expected.
\end{abstract}


\section{Introduction}
\begin{figure}
  \centerline{\psfig{figure=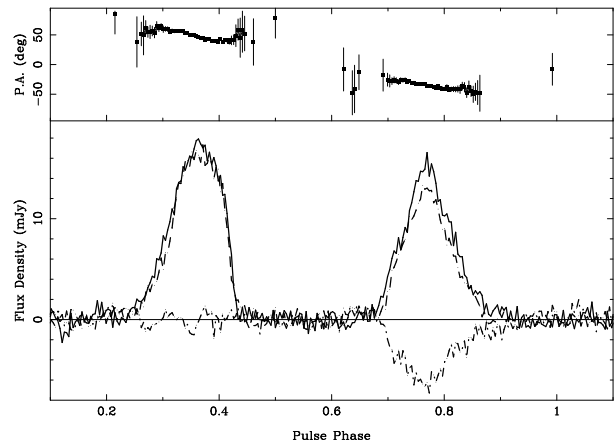,width=8cm,angle=-90}}
  \caption{Mean pulse profile at 1.4~GHz. Position angle is
    shown on top, and the total intensity (solid line), linear (dashed line)
    and circular (dash-dot line) polarisations are shown in the bottom panel.}
  \label{fig:pulseprofile}
\end{figure}

\psr{} was discovered in 1990 in a survey of the Galactic plane
for pulsars \cite{jlm+92} and was subsequently found to be in a
3.4-yr, highly eccentric orbit about a $\sim$10 M$_\odot$ Be star,
\be{} \cite{jml+92}.  It remains the only known radio pulsar with a Be
star companion.  The pulsar has a spin period of $\sim$48~ms, a
characteristic age of 0.33 Myr, and a moderate magnetic field of $3.3
\times 10^{11}$~G.  The dispersion measure (DM) of
146.7~\pow{cm}{-3}pc yields a distance of 4.5 kpc in the model of
Taylor \& Cordes (1993)\nocite{tc93}, however scintillation
\cite{mjsn97} and optical \cite{jml+94} measurements indicate a
distance closer to 1.5 kpc.  The pulse profile, shown in
Figure~\ref{fig:pulseprofile}, has two almost equal intensity peaks
both of which are highly (and almost orthogonally) polarised, with an
average interstellar rotation measure (RM) measured away from
periastron of +21 \rotm{} \cite{mj95}.

Optical observations \cite{jml+94} show that \be{} is of spectral type
B2e, and thus has a mass of $\sim$10 M$_{\odot}$ and a radius R$_{c}\sim$6
R$_{\odot}$. Assuming a pulsar mass of 1.4 M$_{\odot}$, the implied
inclination angle of the binary orbit to the plane of the sky is 36\degr.
The H$\alpha$ emission line shows that the Be star's emission
disc extends to at least 20 R$_{c}$, just inside the pulsar's
orbital radius at periastron.  Timing
measurements have shown that the disc of the Be star is likely to be highly
inclined with respect to the orbital plane \cite{wjm+98}.

The 1997 periastron was observed extensively by
Johnston~et~al.~(1999, 2001)\nocite{jmmc99,jwn+01} and a model for the transient
emission was proposed by Ball et al. (1999)\nocite{bmjs99} following
earlier work by Melatos et al. (1995)\nocite{mjm95}.
In brief, unpulsed emission
began at \p{-22} and lasted until at least \p{+100}
(where $\peri$ denotes the epoch of periastron). The
spectral index of the unpulsed emission was in the range
$-0.5$ to $-0.7$, a value indicative of synchrotron radiation.
Peaks in the light curve at \p{-10} and \p{+20} 
coincide with the pulsar crossing through the emission disc of the Be star.
The model therefore proposed that the synchrotron emission was
generated in the shock between the relativistic pulsar wind and
the outflowing disc of the Be star. Synchrotron losses, set by
the magnetic field strength, then determined the decay time of
the light curve. The decay time of the emission from the post-periastron
disc crossing was substantially shorter than that of the 
pre-periastron decay.

The model predicted a frequency dependence for the end of the
transient phase and verification of this late-time behaviour
was a motivation for the 2000 periastron observations.
Also, the improvements in the correlator and the data reduction
software meant that the pulsar rotation measure (RM)
and dispersion measure (DM) could be
accurately obtained during this periastron passage.
We therefore obtained data over 28 epochs from \p{-46} to \p{+113}
during the 2000 periastron encounter.
In section 2, we describe the observations. In section 3 we
detail the results obtained and discuss their implications
in section 4. 

\section{Observations}  
Observations were made at the Australia Telescope Compact Array (ATCA).
The ATCA is an east-west synthesis telescope located near Narrabri, NSW
and consists of six 22-m antennas on a 6 km track. 
On the ATCA, observations can be made simultaneously at
either 1.4 and 2.4 GHz or 4.8 and 8.4 GHz with a bandwidth of 128 MHz at
each frequency subdivided into 32 spectral channels, and full Stokes
parameters.
The ATCA is also capable of splitting each correlator cycle into 
bins corresponding to different phases of a pulsar's period, and  
in our case the pulse period of $\sim$48~ms was split into 16 bins.

A total of 28 observations between \p{-46} and  \p{+113} were made
using various ATCA configurations. Table~\ref{tab:observations}
lists the relevant details.
One of the ATCA primary flux calibrators, \pria{} (flux densities of
5.5, 5.6, 3.1 and 1.4~Jy at 1.4, 2.5, 4.8 and 8.4~GHz respectively) or
\prib{} (14.9, 11.6, 5.8 and 2.8~Jy), was observed once during each
session for $\sim$5 min. Each pair of bands was switched every
$\sim$20 min, with an observation of the secondary phase calibrator
(\seccal{}; flux density $\sim$1 Jy at all frequencies) for $\sim$3
min leading each integration.
\begin{table}
  \centering
  \begin{tabular}{|ccl|c|c|}
    \multicolumn{3}{|c|}{Start time (UT)} & \multicolumn{1}{c}{Length} & \multicolumn{1}{c}{Array} \\
    \multicolumn{1}{c}{yr} & \multicolumn{1}{c}{mon} & \multicolumn{1}{c}{day} & \multicolumn{1}{c}{(hr)} & \multicolumn{1}{c}{config} \\
    \hline
2000 & Sep & 01.00 & 4 & 6A \\
2000 & Sep & 15.05 & 4 & 6A \\
2000 & Sep & 23.05 & 4 & 6A \\
2000 & Sep & 25.05 & 4 & 6A \\
2000 & Sep & 27.06 & 4 & 6A \\
2000 & Sep & 29.00 & 4 & 6A \\
2000 & Sep & 30.88 & 4 & 6A \\
2000 & Oct & 02.74 & 4 & 6A \\
2000 & Oct & 05.22 & 4 & 6A \\
2000 & Oct & 06.88 & 4 & 6A \\
2000 & Oct & 11.05 & 4 & 6A \\
2000 & Oct & 14.79 & 4 & 6A \\
2000 & Oct & 18.89 & 4 & 6C \\
2000 & Oct & 23.01 & 4 & 6C \\
2000 & Oct & 26.97 & 4 & 6C \\
2000 & Oct & 30.71 & 4 & 6C \\
2000 & Nov & 04.85 & 8 & 6C \\ 
2000 & Nov & 07.79 & 2 & 6C \\ 
2000 & Nov & 10.84 & 4 & 6C \\
2000 & Nov & 15.93 & 4 & 1.5B \\
2000 & Nov & 20.72 & 4 & 1.5B \\
2000 & Nov & 26.72 & 4 & 1.5B \\
2000 & Dec & 08.84 & 4 & 1.5B \\
2000 & Dec & 11.72 & 4 & 1.5B \\
2000 & Dec & 19.86 & 4 & 1.5C \\
2000 & Dec & 26.79 & 5 & 750C \\
2001 & Jan & 08.49 & 5.5 & 750C \\
2001 & Feb & 07.72 & 4 & 6C \\   
    \hline
  \end{tabular}
  \caption{Date, length of observation in hours, and array configuration 
    for all observations of \psrbe{} during the 2000 periastron passage.
    On Nov 7, only 1.4 and 2.4 GHz observations were made. On Dec 8
    time-binning operation was not functional. Periastron occurred
    on Oct 17.3.}
  \label{tab:observations}
\end{table}

Data reduction and analysis were carried out with the \mir{} package
\cite{sk98} using
standard techniques. After flagging bad data, the primary calibrator
was used for flux density and bandpass calibration and the secondary
calibrator was used to solve for antenna gains, phases and
polarisation leakage terms.  The short baselines in the 750C array
typically were completely flagged out at the two lower frequencies,
because of excessive interference.  After calibration, the data
consist of 13 independent frequency channels each 8 MHz wide for each
of the 16 phase bins.

\subsection{Total Flux Density}
To measure the total flux density from the system (i.e. the pulsar plus any
unpulsed emission) we first collapsed the 16 phase bins into one.
Standard techniques of image inversion, cleaning and restoring were
then performed. The flux density and associated errors were 
measured directly from the image with the routine \prog{imfit} with the 
source assumed to be a point source at a fixed position given by the timing data
\cite{wjm+98}.

\subsection{Pulsar Flux Density}
The \prog{uv} binned data were first corrected for dispersion with the
routine \prog{psrfix}.  We assume a constant DM of
146.7~\pow{cm}{-3}pc for all the observations. Although the DM is
known to vary close to periastron, our time resolution of only 3 ms
made the increased time delay largely irrelevant.  We then examined
the pulse profile using and selected those bins which formed the
off-pulse portion of the profile.  The mean of the selected baseline
bins is then subtracted from all the phase bins for every frequency
channel and for each Stokes parameter using \prog{psrbl}.  This has
the effect of removing all sources from the image apart from the
pulsar.  The flux density of the pulsar was then obtained directly
from the visibility data (with \prog{uvflux}) without the need for
further imaging. This process is valid in the absence of sidelobes
from any other source in the primary beam.  The main source of error
in this technique is determining which phase bins are off-pulse. This
is especially true when the pulsar is weak and/or scattered.
Generally we found that the error in the flux density given by the
imaginary component in \prog{uvflux} was equivalent to the value
expected from thermal system noise considerations. Occasionally, a
significantly larger measurement error was determined (indicated with
bold type in Table~\ref{tab:fluxerrs}) which was dominated by phase
calibration errors.
\begin{table*}
   \begin{tabular}{|r|r|*{8}{r}|}
     & &  \multicolumn{8}{c}{Observing Frequency} \\
     \multicolumn{1}{c}{\raisebox{-1ex}[0mm][0mm]{Date}} & \multicolumn{1}{c}{Day}  & \multicolumn{2}{c}{1384 MHz} & 
     \multicolumn{2}{c}{2496 MHz} & \multicolumn{2}{c}{4800 MHz} & 
     \multicolumn{2}{c}{8400 MHz} \\
     & \multicolumn{1}{c}{$(+ \peri)$} & \multicolumn{1}{c}{Pulsar} &
     \multicolumn{1}{c}{Total} & 
     \multicolumn{1}{c}{Pulsar} & \multicolumn{1}{c}{Total} & 
     \multicolumn{1}{c}{Pulsar} & \multicolumn{1}{c}{Total} & 
     \multicolumn{1}{c}{Pulsar} & \multicolumn{1}{c}{Total} \\
     & & \multicolumn{1}{c}{(mJy)} & \multicolumn{1}{c}{(mJy)} & 
     \multicolumn{1}{c}{(mJy)} & \multicolumn{1}{c}{(mJy)} & 
     \multicolumn{1}{c}{(mJy)} & \multicolumn{1}{c}{(mJy)} & 
     \multicolumn{1}{c}{(mJy)} & \multicolumn{1}{c}{(mJy)} \\
     \hline
Sep 01 & $-46.4$ & $5.36$\makebox[8mm][l]{$\pm 0.12$} & $8.00$\makebox[8mm][l]{$\pm 0.73$} & 
$4.77$\makebox[8mm][l]{$\pm 0.19$} & $4.61$\makebox[8mm][l]{$\pm 0.20$} & 
$2.75$\makebox[8mm][l]{$\pm 0.10$} & $2.76$\makebox[8mm][l]{$\pm 0.18$} &
$1.76$\makebox[8mm][l]{$\pm 0.08$} & $1.81$\makebox[8mm][l]{$\pm 0.16$} \\ 
Sep 15 & $-32.4$ & \multicolumn{1}{c}{---} & $3.93$\makebox[8mm][l]{$\pm 0.65$} &
$2.47$\makebox[8mm][l]{$\pm 0.21$} & $5.00$\makebox[8mm][l]{$\pm 0.43$} &
$2.27$\makebox[8mm][l]{$\pm {\bf 1.28}$} & $2.22$\makebox[8mm][l]{$\pm 0.48$} &
$0.61$\makebox[8mm][l]{$\pm {\bf 0.61}$} & $0.78$\makebox[8mm][l]{$\pm 0.32$} \\ 
Sep 23 & $-24.4$ & \multicolumn{1}{c}{---} & $1.64$\makebox[8mm][l]{$\pm 0.37$} & 
\multicolumn{1}{c}{---} & $3.34$\makebox[8mm][l]{$\pm 0.33$} & 
$2.15$\makebox[8mm][l]{$\pm 0.17$} & $2.28$\makebox[8mm][l]{$\pm 0.39$} & 
$0.61$\makebox[8mm][l]{$\pm 0.09$} & $0.64$\makebox[8mm][l]{$\pm 0.19$} \\ 
Sep 25 & $-22.4$ & \multicolumn{1}{c}{---} & $3.78$\makebox[8mm][l]{$\pm 0.56$} & 
$1.38$\makebox[8mm][l]{$\pm 0.20$} & $4.29$\makebox[8mm][l]{$\pm 0.43$} & 
$1.70$\makebox[8mm][l]{$\pm {\bf 0.61}$} & $1.70$\makebox[8mm][l]{$\pm 0.30$} & 
$1.10$\makebox[8mm][l]{$\pm {\bf 0.67}$} & $1.13$\makebox[8mm][l]{$\pm 0.31$} \\ 
Sep 27 & $-20.4$ & \multicolumn{1}{c}{---} & $2.03$\makebox[8mm][l]{$\pm 0.29$} & 
\multicolumn{1}{c}{---} & $1.62$\makebox[8mm][l]{$\pm 0.25$} & 
$0.60$\makebox[8mm][l]{$\pm 0.10$} & $1.43$\makebox[8mm][l]{$\pm 0.34$} & 
$0.44$\makebox[8mm][l]{$\pm 0.09$} & $0.65$\makebox[8mm][l]{$\pm 0.25$} \\ 
Sep 29 & $-18.4$ & $0.29$\makebox[8mm][l]{$\pm 0.13$} & $4.73$\makebox[8mm][l]{$\pm 0.64$} & 
\multicolumn{1}{c}{---} & $5.36$\makebox[8mm][l]{$\pm 0.34$} & 
$0.57$\makebox[8mm][l]{$\pm 0.14$} & $3.26$\makebox[8mm][l]{$\pm 0.40$} & 
$0.22$\makebox[8mm][l]{$\pm 0.09$} & $1.58$\makebox[8mm][l]{$\pm 0.40$} \\ 
Sep 30 & $-16.5$ & \multicolumn{1}{c}{---} & $6.44$\makebox[8mm][l]{$\pm 1.06$} & 
\multicolumn{1}{c}{---} & $6.86$\makebox[8mm][l]{$\pm 0.31$} & 
\multicolumn{1}{c}{---} & $4.62$\makebox[8mm][l]{$\pm 0.38$} & 
\multicolumn{1}{c}{---} & $2.86$\makebox[8mm][l]{$\pm 0.42$} \\ 
Oct 02 & $-14.7$ & \multicolumn{1}{c}{---} & $17.60$\makebox[8mm][l]{$\pm 1.91$} & 
\multicolumn{1}{c}{---} & $13.28$\makebox[8mm][l]{$\pm 1.29$} & 
\multicolumn{1}{c}{---} & $9.17$\makebox[8mm][l]{$\pm 2.41$} & 
\multicolumn{1}{c}{---} & $3.18$\makebox[8mm][l]{$\pm 1.85$} \\ 
Oct 05 & $-12.2$ & \multicolumn{1}{c}{---} & $27.39$\makebox[8mm][l]{$\pm 2.62$} & 
\multicolumn{1}{c}{---} & $23.27$\makebox[8mm][l]{$\pm 1.57$} & 
\multicolumn{1}{c}{---} & $12.06$\makebox[8mm][l]{$\pm 3.71$} & 
\multicolumn{1}{c}{---} & $3.09$\makebox[8mm][l]{$\pm 2.20$} \\ 
Oct 06 & $-10.5$ & \multicolumn{1}{c}{---} & $33.37$\makebox[8mm][l]{$\pm 1.68$} & 
\multicolumn{1}{c}{---} & $29.61$\makebox[8mm][l]{$\pm 0.78$} & 
\multicolumn{1}{c}{---} & $21.95$\makebox[8mm][l]{$\pm 1.16$} & 
\multicolumn{1}{c}{---} & $13.68$\makebox[8mm][l]{$\pm 1.11$} \\ 
Oct 11 & $-6.4$ & \multicolumn{1}{c}{---} & $39.64$\makebox[8mm][l]{$\pm 2.91$} & 
\multicolumn{1}{c}{---} & $29.83$\makebox[8mm][l]{$\pm 2.01$} & 
\multicolumn{1}{c}{---} & $13.93$\makebox[8mm][l]{$\pm 3.33$} & 
\multicolumn{1}{c}{---} & $3.27$\makebox[8mm][l]{$\pm 2.05$} \\ 
Oct 14 & $-2.6$ & \multicolumn{1}{c}{---} & $29.16$\makebox[8mm][l]{$\pm 1.70$} & 
\multicolumn{1}{c}{---} & $23.70$\makebox[8mm][l]{$\pm 0.84$} & 
\multicolumn{1}{c}{---} & $16.17$\makebox[8mm][l]{$\pm 1.97$} & 
\multicolumn{1}{c}{---} & $9.18$\makebox[8mm][l]{$\pm 1.98$} \\ 
Oct 18 & $1.5$ & \multicolumn{1}{c}{---} & $21.50$\makebox[8mm][l]{$\pm 1.59$} & 
\multicolumn{1}{c}{---} & $17.36$\makebox[8mm][l]{$\pm 0.54$} & 
\multicolumn{1}{c}{---} & $11.75$\makebox[8mm][l]{$\pm 0.99$} & 
\multicolumn{1}{c}{---} & $7.85$\makebox[8mm][l]{$\pm 0.99$} \\ 
Oct 23 & $5.6$ & \multicolumn{1}{c}{---} & $21.72$\makebox[8mm][l]{$\pm 1.24$} & 
\multicolumn{1}{c}{---} & $16.93$\makebox[8mm][l]{$\pm 1.07$} & 
\multicolumn{1}{c}{---} & $10.15$\makebox[8mm][l]{$\pm 2.52$} & 
\multicolumn{1}{c}{---} & $3.70$\makebox[8mm][l]{$\pm 2.21$} \\ 
Oct 26 & $9.6$ & \multicolumn{1}{c}{---} & $22.57$\makebox[8mm][l]{$\pm 1.62$} & 
\multicolumn{1}{c}{---} & $15.88$\makebox[8mm][l]{$\pm 1.19$} & 
\multicolumn{1}{c}{---} & $6.79$\makebox[8mm][l]{$\pm 1.54$} & 
\multicolumn{1}{c}{---} & $1.22$\makebox[8mm][l]{$\pm 0.50$} \\ 
Oct 30 & $13.3$ & \multicolumn{1}{c}{---} & $22.55$\makebox[8mm][l]{$\pm 2.17$} & 
\multicolumn{1}{c}{---} & $17.50$\makebox[8mm][l]{$\pm 1.07$} & 
\multicolumn{1}{c}{---} & $12.35$\makebox[8mm][l]{$\pm 1.39$} & 
\multicolumn{1}{c}{---} & $6.13$\makebox[8mm][l]{$\pm 1.59$} \\ 
Nov 04 & $18.5$ & \multicolumn{1}{c}{---} & $29.88$\makebox[8mm][l]{$\pm 1.46$} & 
\multicolumn{1}{c}{---} & $23.22$\makebox[8mm][l]{$\pm 0.92$} & 
$0.39$\makebox[8mm][l]{$\pm 0.06$} & $11.02$\makebox[8mm][l]{$\pm 1.42$} & 
$0.36$\makebox[8mm][l]{$\pm 0.05$} & $2.46$\makebox[8mm][l]{$\pm 0.76$} \\ 
Nov 07 & $21.3$ & $1.64$\makebox[8mm][l]{$\pm 0.12$} & $35.66$\makebox[8mm][l]{$\pm 2.54$} & 
$3.31$\makebox[8mm][l]{$\pm 0.21$} & $34.65$\makebox[8mm][l]{$\pm 1.32$} 
 \\
Nov 10 & $24.4$ & $2.97$\makebox[8mm][l]{$\pm 0.17$} & $36.95$\makebox[8mm][l]{$\pm 2.27$} & 
$5.12$\makebox[8mm][l]{$\pm {\bf 1.03}$} & $34.60$\makebox[8mm][l]{$\pm 1.46$} & 
$4.33$\makebox[8mm][l]{$\pm 0.18$} & $22.95$\makebox[8mm][l]{$\pm 2.01$} & 
$1.59$\makebox[8mm][l]{$\pm {\bf 0.58}$} & $13.40$\makebox[8mm][l]{$\pm 1.55$} \\ 
Nov 15 & $29.5$ & $3.75$\makebox[8mm][l]{$\pm 0.18$} & $32.04$\makebox[8mm][l]{$\pm 3.15$} & 
$3.08$\makebox[8mm][l]{$\pm 0.12$} & $24.62$\makebox[8mm][l]{$\pm 1.15$} & 
$1.89$\makebox[8mm][l]{$\pm {\bf 0.27}$} & $13.81$\makebox[8mm][l]{$\pm 1.86$} & 
$1.48$\makebox[8mm][l]{$\pm 0.13$} & $7.06$\makebox[8mm][l]{$\pm 1.72$} \\ 
Nov 20 & $34.3$ & $3.33$\makebox[8mm][l]{$\pm 0.13$} & $26.69$\makebox[8mm][l]{$\pm 2.70$} & 
$3.61$\makebox[8mm][l]{$\pm 0.17$} & $23.61$\makebox[8mm][l]{$\pm 0.80$} & 
$2.94$\makebox[8mm][l]{$\pm 0.09$} & $17.27$\makebox[8mm][l]{$\pm 1.06$} & 
$0.86$\makebox[8mm][l]{$\pm 0.08$} & $9.62$\makebox[8mm][l]{$\pm 0.99$} \\ 
Nov 26 & $40.3$ & $3.69$\makebox[8mm][l]{$\pm 0.11$} & $20.73$\makebox[8mm][l]{$\pm 1.73$} & 
$3.83$\makebox[8mm][l]{$\pm 0.17$} & $18.92$\makebox[8mm][l]{$\pm 0.64$} & 
$1.63$\makebox[8mm][l]{$\pm 0.08$} & $11.06$\makebox[8mm][l]{$\pm 0.77$} & 
$1.60$\makebox[8mm][l]{$\pm {\bf 0.45}$} & $6.88$\makebox[8mm][l]{$\pm 0.85$} \\ 
Dec 08 & $52.3$ & & $18.47$\makebox[8mm][l]{$\pm 2.28$} & 
                  & $13.36$\makebox[8mm][l]{$\pm 0.72$} & 
                  & $7.83$\makebox[8mm][l]{$\pm 0.77$} & 
                  & $4.02$\makebox[8mm][l]{$\pm 0.73$} \\ 
Dec 11 & $55.3$ & $4.05$\makebox[8mm][l]{$\pm 0.15$} & $15.47$\makebox[8mm][l]{$\pm 1.72$} & 
$3.56$\makebox[8mm][l]{$\pm 0.19$} & $11.70$\makebox[8mm][l]{$\pm 0.38$} & 
$1.11$\makebox[8mm][l]{$\pm 0.12$} & $5.43$\makebox[8mm][l]{$\pm 0.32$} & 
$1.65$\makebox[8mm][l]{$\pm 0.10$} & $3.80$\makebox[8mm][l]{$\pm 0.33$} \\ 
Dec 19 & $63.4$ & $4.41$\makebox[8mm][l]{$\pm 0.20$} & $15.58$\makebox[8mm][l]{$\pm 1.62$} & 
$3.64$\makebox[8mm][l]{$\pm 0.20$} & $10.30$\makebox[8mm][l]{$\pm 0.71$} & 
$2.38$\makebox[8mm][l]{$\pm {\bf 0.49}$} & $5.07$\makebox[8mm][l]{$\pm 1.35$} & 
$0.57$\makebox[8mm][l]{$\pm {\bf 0.27}$} & $1.39$\makebox[8mm][l]{$\pm 0.61$} \\ 
Dec 26 & $70.4$ & $5.09$\makebox[8mm][l]{$\pm 0.18$} & $10.25$\makebox[8mm][l]{$\pm 1.84$} & 
$4.11$\makebox[8mm][l]{$\pm 0.12$} & $7.47$\makebox[8mm][l]{$\pm 0.55$} & 
$1.92$\makebox[8mm][l]{$\pm 0.08$} & $5.19$\makebox[8mm][l]{$\pm 0.55$} & 
$1.07$\makebox[8mm][l]{$\pm 0.06$} & $0.75$\makebox[8mm][l]{$\pm 0.41$} \\ 
Jan 08 & $83.1$ & $5.28$\makebox[8mm][l]{$\pm {\bf 1.14}$} & $13.52$\makebox[8mm][l]{$\pm 3.46$} & 
$4.76$\makebox[8mm][l]{$\pm {\bf 0.84}$} & $10.38$\makebox[8mm][l]{$\pm 0.60$} & 
$1.85$\makebox[8mm][l]{$\pm 0.09$} & $5.09$\makebox[8mm][l]{$\pm 0.41$} & 
$0.48$\makebox[8mm][l]{$\pm 0.06$} & $2.56$\makebox[8mm][l]{$\pm 0.38$} \\ 
Feb 07 & $113.3$ & $3.60$\makebox[8mm][l]{$\pm {\bf 0.99}$} & $10.85$\makebox[8mm][l]{$\pm 0.86$} & 
$3.39$\makebox[8mm][l]{$\pm {\bf 1.24}$} & $7.97$\makebox[8mm][l]{$\pm 0.63$} & 
$3.87$\makebox[8mm][l]{$\pm {\bf 2.94}$} & $6.46$\makebox[8mm][l]{$\pm 2.08$} & 
$0.74$\makebox[8mm][l]{$\pm {\bf 1.35}$} & $2.19$\makebox[8mm][l]{$\pm 1.31$} \\ 
        \hline
   \end{tabular}
   \caption{Flux densities for pulsed emission and total emission (pulsed
     plus unpulsed) for all 29 observations with the ATCA.
     Flux densities and errors for continuum and pulsar are from \prog{imfit}.
     Flux densities and errors for pulsar only are from \prog{uvflux}.
     Pulsar flux densities are omitted when no significant pulses could
     be seen.}
   \label{tab:fluxerrs}
\end{table*}

\subsection{Dispersion Measure}
Even with our relatively crude time resolution of $\sim$3 ms per bin
we are able to measure the DM across the band at 1.4 GHz with
reasonable accuracy. To achieve this we determined the centroid of
each of the two pulse components for each of the 13 frequency
channels.  The time delay in seconds for a given channel, $c$,
relative to the first channel is given by (see e.g. Manchester \&
Taylor 1977\nocite{mt77})
\beq
t_c \,\,\, = \,\,\, 4149.94 \,\,\, {\rm DM} \,\,\,
   \left(\frac{1}{\nu^{2}_{c}} - \frac{1}{\nu^{2}_{1}}\right)
\eeq
where $\nu$ is the frequency of the channel in MHz.  The DM can
therefore be obtained by a straight-line fit to a plot of dispersion
delay versus wavelength squared. The error bar on the DM is 
about one order of magnitude higher than from `conventional'
pulsar timing \cite{jwn+01}.

\subsection{Circular Polarisation}
As the natural modes of the wind of the Be star and the
interstellar plasma are circular,
circularly polarised flux is not affected by propagation.
We therefore used the measurement of
the circular polarisation under each pulse component independently
to verify our polarisation calibration.
As before, the flux in Stokes $V$ for two separate phase bin 
ranges corresponding to the two pulsar components was
obtained from the visibility data via \prog{uvflux}.

\subsection{Linear Polarisation and RM}
\prog{uvflux} was used to measure Stokes $Q$ and $U$ in each
channel for each phase bin and/or combinations of channels and
phase bins. In practice, in order to boost signal to noise,
we combine the phase bins together which correspond to each pulse 
component as Stokes $Q$ and $U$ do not vary rapidly through the pulse
(see Fig 1).
The position angle, $\phi_\lambda$, of each pulse component 
at a given observing wavelength, $\lambda$, can be then be computed through
\beq
  \phi_\lambda =0.5 \times \tan^{-1}(U_\lambda/Q_\lambda).
  \label{eqn:polangle}
\eeq
The error on the position angle is given by
\beq
 \Delta\phi = 0.5 \epsilon / \sqrt{Q^2 + U^2}
\eeq
when both the $Q$ and $U$ measurements have the same rms error $\epsilon$.
Note that as the linear polarisation has a positive definite bias,
this must be subtracted before the error is computed.
The position angle is related to the RM by
\beq
  \phi_\nu = {\rm RM}\,\lambda^2
  \label{eqn:rmphi}
\eeq
Hence, plotting position angle against $\lambda^2$ should yield
a straight line with slope RM. This procedure was carried
out for each of the two components independently.

\section{Results} 
The flux densities and their associated errors
from the observations listed in
Table~\ref{tab:observations} are given in Table~\ref{tab:fluxerrs}.
The table first lists
the date of the observation, followed by the time from periastron in
days. The rest of the table contains the flux
density measurements of the pulsar (when detectable) 
and the total flux density (pulsed plus unpulsed) obtained at each of the
four frequencies using the techniques described in the previous section.
\begin{figure}
  \centerline{\psfig{figure=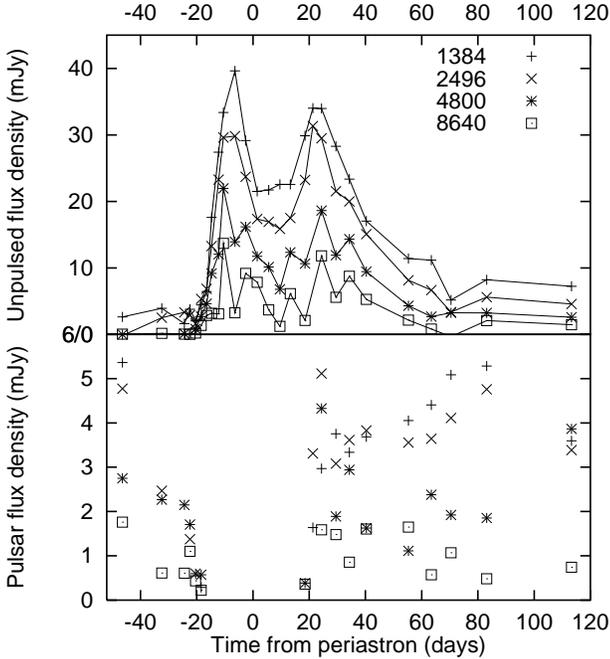,width=8cm,angle=-90}}
  \caption{Light curves for the unpulsed emission (top) and the
           pulsar (bottom) from the 2000 periastron data. For clarity 
           the error bars have been omitted.}
  \label{fig:lightcurve}
\end{figure}

\subsection{Pulsed Emission}
The pulsar fluxes are shown in the lower panel of Figure~\ref{fig:lightcurve}.
Before periastron, the pulsar was detected at all frequencies at
\p{-46}. At \p{-32} the pulsar was scattered at 2.5~GHz and not detected at
1.4~GHz, at \p{-24} scattering was evident at 4.8~GHz and the pulsar
not detected at the two lower frequencies. However, at \p{-22}
the pulsar was again detected at 2.5~GHz but was scattered at
that frequency two days later.
At \p{-18} the evidence for pulsed emission at any frequency is marginal
although it cannot be ruled out. Following this date there is no
further detection of pulsed emission until \p{+18}.
Post-periastron, there is a clear detection at the two higher frequencies
at \p{+18} and shortly after the end of the ATCA observation the pulsar
was also detected at Parkes at 1.3~GHz. Following this, the pulsar
was detectable again at all frequencies.

It seems likely therefore that scattering dominates the 
detectability of the pulsar
for $\sim$20 days leading up to \p{-16}.
From this date until \p{+18} the pulsar is behind the emission disc
with respect to the observer
and the pulsed emission is quenched by the high optical depth.
The pulsar then appears to emerge from the disc rather abruptly;
the scattering is marginal although optical depth effects are
still visible at 1.4~GHz until $\p{+25}$.
Unfortunately, the poor time resolution of our data precludes
measurement of the scattering parameters.

In 1997 the pulsar was detected with the ATCA until \p{-21} at all
frequencies and at Parkes until \p{-18}. Post-periastron it was
detected from \p{+16} onwards. The effects of the wind on the pulsed emission
therefore occurred earlier and lasted longer in 2000 than in 1997.

\subsection{Unpulsed emission}
\begin{figure}
  \centerline{\psfig{figure=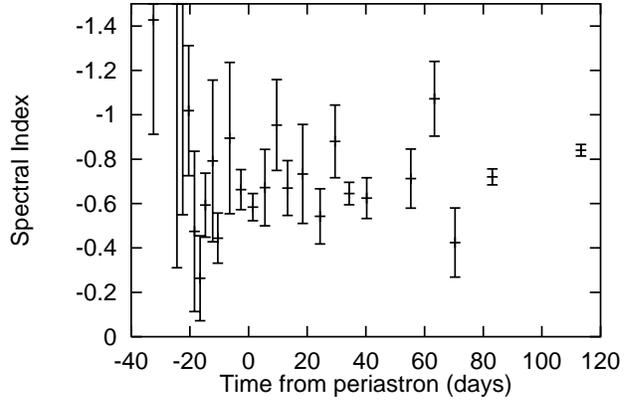,width=8.1cm}}
  \caption{Spectral indices for the unpulsed emission.}
  \label{fig:si2000}
\end{figure}
The top panel of Figure~\ref{fig:lightcurve} shows the 
light curve for the unpulsed emission.
Unpulsed emission appears to be present at 1.4 GHz during
the first observation at $\sim \p{-46}$ and remains roughly
constant until $\sim \p{-20}$. The flux
density then rapidly increases (to 40~mJy at 1.4~GHz) towards the first
peak at $\sim \p{-6}$. A decrease with a similar slope follows,
between $\sim \p{-6}$ and \p{+2}. The flux density plateaus (at $\sim
21$~mJy at 1.4~GHz) at all frequencies between \p{+1} and \p{+13}, and
then rises again to a peak at $\sim \p{+22}$. The rise to the second
peak is slower than the first, and the subsequent decay is rather shallow.
\p{+70} shows a sudden drop in the flux density, subsequently however
the unpulsed emission remains clearly present at the two lower
frequencies during our final
observation at \p{+113}. This is similar to data taken in 1994
which shows unpulsed emission present at 0.84 GHz until at 
least \p{+132} \cite{jmmc99}.
The data do show, however, that unpulsed emission is not
present at 8.4 GHz from \p{+63} onwards and
has also vanished at 4.8 GHz by \p{+113}.

\subsection{Spectral indices}
\begin{table*}
  \begin{tabular}{|rr|c|ccccc}
    \multicolumn{1}{c|}{Date} & \multicolumn{1}{c|}{(+\peri)} & 
    \multicolumn{1}{c|}{Time} &
    \multicolumn{4}{c|}{Rotation Measure (rad~m$^{-2})$} & 
    \multicolumn{1}{c}{DM} \\
    & & \multicolumn{1}{c|}{UT} & \multicolumn{1}{c|}{1384 MHz} &
        \multicolumn{1}{c|}{2496 MHz} & 
        \multicolumn{1}{c|}{4800 MHz} & \multicolumn{1}{c|}{8640 MHz} 
    & \multicolumn{1}{c}{cm$^{-3}$pc}\\
    \hline
    Sep 01 & $-46.4$ & 
    00:45 -- 01:05 & & & $-14800$\makebox[8mm][l]{$\pm 1800$} \\
    & & 01:35 -- 01:55 & & & $-6900$\makebox[8mm][l]{$\pm 5000$} \\
    & & 02:25 -- 02:45 & & & $-12500$\makebox[8mm][l]{$\pm 1400$} \\
    & & 00:45 -- 03:45 & & & & $-15200$\makebox[8mm][l]{$\pm 3500$} &
    148.7\makebox[8mm][l]{$\pm 1.5$}\\
    Nov 07 & 21.3 & & & $-5100$\makebox[8mm][l]{$\pm 360$} \\
    Nov 10 & 24.4 & & $-450$\makebox[8mm][l]{$\pm 60$} & 
    $-460$\makebox[8mm][l]{$\pm 180$} & & &
    145.7\makebox[8mm][l]{$\pm 2.9$}\\
    Nov 15 & 29.5 &
    22:30 -- 22:50 & & & $+7700$\makebox[8mm][l]{$\pm 2100$} & &
    148.9\makebox[8mm][l]{$\pm 2.9$}\\
    Nov 20 & 34.3 & & $-360$\makebox[8mm][l]{$\pm 5$} &
    $-350$\makebox[8mm][l]{$\pm 120$} & & &
    144.1\makebox[8mm][l]{$\pm 3.7$}\\
    Nov 26 & 40.3 & & $-675$\makebox[8mm][l]{$\pm 30$} &
    $-750$\makebox[8mm][l]{$\pm 100$} & & &
    144.4\makebox[8mm][l]{$\pm 2.3$}\\
    Dec 11 & 55.3 & & $-101$\makebox[8mm][l]{$\pm 9$} & 
    $-75$\makebox[8mm][l]{$\pm 30$} & & &
    148.4\makebox[8mm][l]{$\pm 1.6$}\\
    Dec 19 & 63.4 &
    21:05 -- 21:30 & & & $-7700$\makebox[8mm][l]{$\pm 2400$} & &
    146.9\makebox[8mm][l]{$\pm 2.4$}\\
    Dec 26 & 70.4 & 
    19:25 -- 21:00 & $-94$\makebox[8mm][l]{$\pm 14$} \\
    & & 21:00 -- 23:30 & $-50$\makebox[8mm][l]{$\pm 7$} & 
    $-102$\makebox[8mm][l]{$\pm 50$} & & &
    147.8\makebox[8mm][l]{$\pm 2.0$}\\
    Jan 08 & 83.1 & & $-49$\makebox[8mm][l]{$\pm 9$} & 
    $-24$\makebox[8mm][l]{$\pm 35$} & & &
    147.6\makebox[8mm][l]{$\pm 1.6$}\\
    Feb 07 & 113.3 & & $13$\makebox[8mm][l]{$\pm 11$} & 
    $130$\makebox[8mm][l]{$\pm 80$} & & &
    145.1\makebox[8mm][l]{$\pm 2.5$}\\
    \hline
  \end{tabular}
\caption{RM and DM of the pulsar at various epochs
  averaged over the two pulse components. The errors are 1~$\sigma$.
  Where no time range is given the entire observation was used.}
\label{tab:rm}
\end{table*}
%\begin{figure}
%  \centerline{\psfig{figure=fig4.ps,width=8.1cm}}
%  \caption{Percentage circular polarisations for the first (left panel)
%      and second (right panel) pulse component at each of the four
%      frequencies (1.4 GHz at the top). Note the different scales
%      for each component.
%      The mean value is shown as a horizontal line.}
%  \label{fig:circ}
%\end{figure}
The flux density at a frequency $\nu$
is given by $S_\nu=C \nu^{\alpha}$
where $\alpha$ denotes the spectral index. We measure the spectral
index for the non-pulsed emission using flux densities 
obtained at each of our four observing frequencies.
We apply a simple linear fit to log $S$ - log $\nu$ space and assume
symmetrical errors in log $S$.
The spectral indices are shown in Figure~\ref{fig:si2000}
and are consistent with a value of $\sim -0.7$ throughout.
There is no evidence for optical depths effects, at least at
frequencies above 1 GHz. This is similar to the behaviour in
1997 \cite{jmmc99}.

We do not attempt to measure the spectral index of the pulsed emission.
At 8.4 GHz, diffractive scintillation significantly affects the
observed flux density. The scintillation bandwidth of $\sim$70~MHz
and timescale of $\sim$10~min \cite{mjsn97} match well with
our observing bandwidth of $\sim$120~MHz and time of 20~min.
Furthermore, there is clear evidence for optical depth effects at
low frequency post-periastron. In particular, the flux density at
1.4 GHz appears to increase steadily from \p{+21} to \p{+70}
whereas this is not the case at the two higher frequencies.
\begin{figure}
  \centerline{\psfig{figure=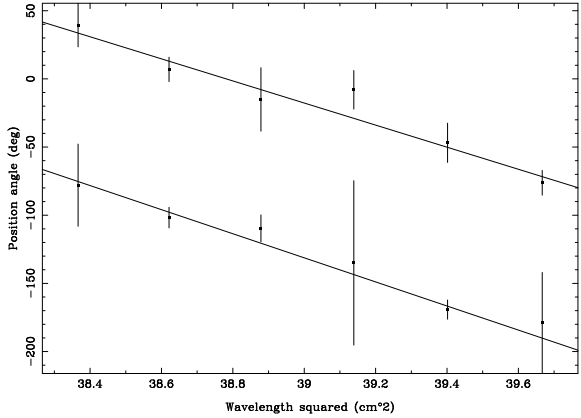,width=7.6cm,angle=-90}}
  \caption{Position angle versus $\lambda^{2}$ across the
    4.8 GHz band on September 1 for the first (top) and second
    (bottom) components. The position angle separation between the 
    components is near 110\deg{}, as expected.
    The straight lines show the fit to the
    data from which the RM can be derived.}
  \label{fig:rmsep01}
\end{figure}

\subsection{Polarisation and Dispersion Measure}
\begin{figure*}
  \centerline{\psfig{figure=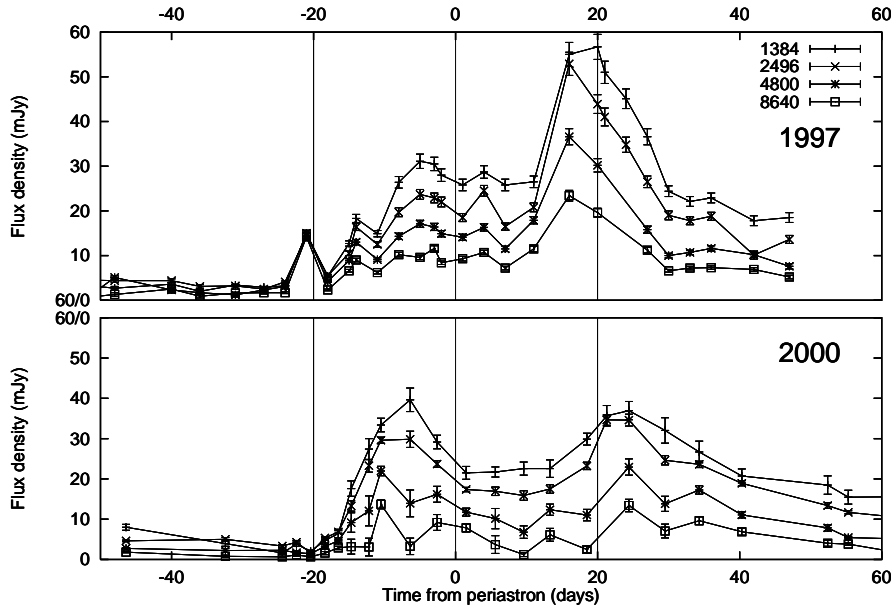,width=11.7cm,angle=-90}}
  \caption{Light curves from the 1997 and 2000 data (pulsed and unpulsed
    flux combined) on the same scale. The vertical lines at 
    $\peri \pm 20$ delineate the approximate eclipse of the pulsar.}
  \label{fig:1997-2000light}
\end{figure*}
%The fractional circular polarisation
%for the first and second pulse components, 
%are presented in Fig~\ref{fig:circ}.
The circular polarisation was obtained for each pulse component for
each observation, with errors of order $\sim 5 \%$ at the lower
frequencies, and $\sim 10 \%$ at the higher frequencies due to the
higher system temperature and lower flux density at these
frequencies. The data are consistent with a mean value in the range
$-6 \%$ to $-10 \%$ for the first component, and between $-28 \%$ and
$-30 \%$ for the second component for all frequencies, in agreement
with observations made at Parkes (see also
Fig~\ref{fig:pulseprofile}).  We are thus confident that our
polarisation calibration and analysis techniques are correct.

We attempted to obtain RMs and DMs for
all days when the pulsar was not eclipsed using the techniques 
described in Section 2. These are presented in Table~\ref{tab:rm}.
The RM values are weighted averages of each of the two pulse 
components. In all cases
the RMs agreed within the errors.
Note that in general for RMs \lapp 2000 \rotm{}, only
the two lower frequencies will show a significant position angle
change across the bandpass whereas for RMs \gapp 5000 \rotm{}
depolarisation will occur even across a single channel at 2.4 GHz.

Prior to periastron, \p{-46} was the only occasion on which we
were able to derive an RM from the pulsar. At 8.4 GHz, the value we derived
is $-15200$ \rotm{}. As detailed in Section 2, the data are
obtained in 20 min sections at the high frequencies followed by
20 min at the low frequencies. We therefore divided the 4.8 GHz
data into four 20 min sections and derived the RMs independently.
Figure \ref{fig:rmsep01} shows the position angles obtained
across the 4.8 GHz band during the third observation on \p{-46} (Sep 1)
and the fitted slope.
The RMs measured on this date are extremely high,
the highest astrophysical RM measurement as far as we are aware.
There is also marginal evidence for a varying RM throughout
the observing period.
We presume that the RM increased even further after this date
ensuring depolarisation even at 8.4 GHz on Sep 15 (\p{-32}).

On \p{+18} there is a marginal detection of the pulsar but no RM or DM
can be obtained. On \p{+21}, only the two lowest frequencies were
observed but polarisation is clearly seen at 2.4 GHz and
we derive an RM of $-5100$~\rotm{}. On \p{+24}, the RM has
dropped significantly, however, on \p{+29} the RM is large
and positive but can only be detected
in the first 20 min integration at 4.8 GHz as detailed in the table.
We note that this is the only positive value of RM obtained.
A similar result was obtained during the 1994 periastron, where the RMs
were consistently negative both before and after periastron.
Finally, as late as \p{+63} the pulsar appears depolarised at all
but 8.4 GHz. During the first 20 min observation at 4.8 GHz we
measured an RM of $-7700$~\rotm{}.

There is only marginal evidence for a variable DM with epoch, with the
errors quoted in the table being much larger than the typical 0.2 --
0.4~cm$^{-3}$pc obtained using pulsar timing \cite{jml+96}.  The mean
of the values quoted is 146.8 cm$^{-3}$pc, almost exactly the result
obtained from timing.  We note that these results are consistent with
the values obtained during the 1994 and 1997 periastra. In 1994, the
$\Delta$DM (i.e.  the DM with respect to 146.8 cm$^{-3}$pc) was
3.7~cm$^{-3}$pc by \p{-28} and no change in DM was seen
post-periastron \cite{jml+96}.  In 1997, the $\Delta$DM was
1.5~cm$^{-3}$pc on \p{-48} and 4~cm$^{-3}$pc on
\p{-42}. An upper limit of 0.2~cm$^{-3}$pc to the $\Delta$DM was
obtained post-periastron \cite{jwn+01}.

From the current data we therefore set an upper limit to $\Delta$DM
on \p{-46} of 3.0 cm$^{-3}$pc and on \p{+29}
of 5.8 cm$^{-3}$pc (i.e. twice the error bars), although from the
above discussion the $\Delta$DM is likely to be much smaller
on the latter date.
The magnetic field parallel to the line of sight, $B_{\Vert}$ in
$\mu$G, can be obtained with knowledge of the RM and $\Delta$DM by
\beq
B_{\Vert} \,\,\, = 1.232 \,\,\, \frac{{\rm RM}}{{\rm \Delta{}DM}}
\eeq
where RM and $\Delta$DM are in conventional units.
The lower limits of $B_{\Vert}$ on \p{-46} (Sep 1)
and \p{+29} (Nov 15) are therefore
$6.2$~mG and $1.6$~mG respectively.

\section{Discussion}
A comparison of the light curves of 1997 and 2000 appears in
Figure~\ref{fig:1997-2000light}. The same basic behaviour is
seen for both sets of observations. There is a clear peak
before and after periastron with a plateau in between. The
transient emission also persists until late times in both.
However, the initial `blip' in the light curve seen at \p{-21} in
1997 was not repeated in 2000. Johnston et al. (2001) identified
this as a transient event possibly associated with the pulsar
`splashing' into the disc.
Also, in 2000, the first peak occurs earlier
and the second peak later than in 1997. This is consistent with
the longer lasting eclipse of the pulsed emission.
The other striking feature is that the second peak is significantly
lower in 2000 than in 1997. 
In broad terms, therefore, the model proposed after the 1997 periastron
is repeated again. The pulsar enters the disc of the Be star
near \p{-20}, electrons are accelerated in the interface between
the pulsar wind and the disc material. This occurs over the few
days that the pulsar takes to cross the disc. When the pulsar leaves
the disc, the wind bubble remains behind and decays through
synchrotron losses. During the second crossing of the disc near
\p{+20} the same process happens again causing the second peak seen
in the light curve.
\begin{figure}
  \centerline{\psfig{figure=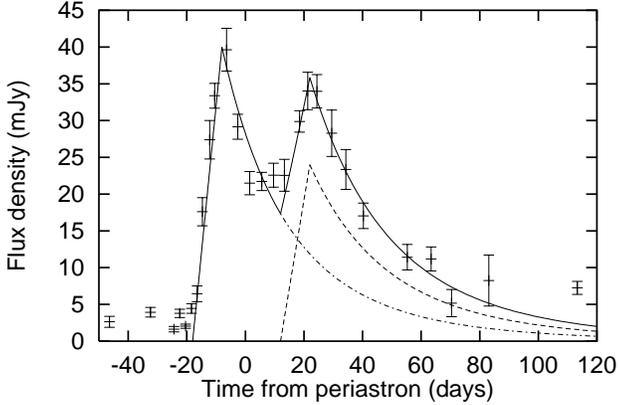,width=8.1cm}}
  \caption{Flux density of the unpulsed emission at 1.4 GHz. Superposed
    is the solution to the model described in the text. The model includes
    a rise time of 10 days and adiabatic expansion for each of the
    synchrotron bubbles created in the disc crossings. The dashed lines
    indicate the light curve for each bubble individually. The solid
    line is the sum of both light curves.}
  \label{fig:addplateau}
\end{figure}

In Fig~\ref{fig:addplateau} we again show the light curve at 1.4 GHz.
The decay of the second peak is clearly not linear, but follows a more
power-law like behaviour. This is the expected form of the flux
density for adiabatic expansion of a synchrotron bubble.
We can write the synchrotron flux density, $S_\nu$, from an expanding bubble as
\beq
S_\nu \propto L^{2\alpha -2} \,\,\, B^{1-\alpha} \,\,\, \nu^{\alpha}
\label{eqn:st1}
\eeq
where $L$ is the radius of the bubble, $B$ is the magnetic field,
and $\nu$ and $\alpha$ are the frequency and spectral index respectively.
Imagine the synchrotron bubble is formed at an initial distance $R_0$
from the Be star and with an initial radius $L_0$. We then assume
that the bubble moves outwards in the disc of the Be star with
a velocity $V$. Hence
\beq
R(t) = R_0 \,\,\, + \,\,\, V \, (t-t_0)
\eeq
where $t_0$ is the start time of the adiabatic phase.  The change in
$R$ is evidently $R(t)/R_0$. We assume that the density in the
disc follows a power-law with distance from the star i.e. $\rho(R) = C
R^{-n}$. It then follows that
\beq
\frac{\rho(t)}{\rho_0} = \left(\frac{R_0 \,\,\, + \,\,\, V \, (t-t_0)}{R_0}\right)^{-n}
\eeq
As the size of the bubble is dictated by the ram pressure between the
pulsar wind and the disc material then $L \propto \rho^{-1/2}$ and so
\beq
\frac{L}{L_0} = \left(\frac{R_0 \,\,\, + \,\,\, V \, (t-t_0)}{R_0}\right)^{n/2}
\eeq
The magnetic field in the disc is radial at the distances relevant here
\cite{mjm95} and so
\beq
\frac{B}{B_0} = \left(\frac{R_0 \,\,\, + \,\,\, V \, (t-t_0)}{R_0}\right)^{-1}
\eeq
Substituting these last two equations into equation~\ref{eqn:st1} above
we finally obtain
\beq
S_\nu = S_0 \,\,\, \left(\frac{R_0 \,\,\, + \,\,\, V \, (t-t_0)}{R_0}\right)^{(\alpha -1)(1+n)} \,\,\, \nu^{\alpha}
\label{eqn:st2}
\eeq

We model the data as follows.
We set $\alpha = -0.6$ from the observations and
assume $n=3$ \cite{wvt+91}. We assume that $t_0$ for the first
encounter is \p{-8} and for the second is \p{+21}. The distance from
the Be star at these epochs are $34~R_*$ and $42~R_*$ respectively.
The interaction between the pulsar wind and the Be star's disc
continues for $\sim$10 days. During this time electrons are injected
continuously and a linear rise in the flux density
provides a good fit to the data.
We then fit the entire data set using the functional form of
equation~\ref{eqn:st2} and with $S_0$ for both encounters and $V$
as free parameters.
The solution yields $S_0$ of 40 and 24~mJy respectively (at 1.4 GHz) and
$V = 12$~km~s$^{-1}$ and this solution is shown in 
Fig~\ref{fig:addplateau}. The ratio of the peak fluxes is in agreement
with the relative distance from the Be star of the two start epochs.
We note that although the fit is generally good
it falls well below the measured flux density of the last data 
point at \p{+113}.
It is also evident that some unpulsed emission is present as early
as \p{-46} at least at low frequencies. There therefore appears to be
some low-level emission unrelated to the disc crossings.
This emission could arise with the interaction between the pulsar wind
and dense clumps in the (more tenuous) Be star wind.
We repeated the fitting process but in addition we had a constant offset
of 4~mJy throughout the entire observing period to represent the
low-level emission. This naturally provides an improved fit to both
very early and very late times. However, $V$ is largely unchanged at
15~km~s$^{-1}$.

There is a strong similarity between the system under discussion here
and the \lsi\ binary. There, an unseen neutron star orbits a Be star
in a highly eccentric orbit with a period of only 26 days.
Radio outbursts show a steep rise followed by a power-law like decay.
Paredes et al. (1991)\nocite{pmes91}, in a more complete treatment,
modelled the radio light curve of \lsi\
in a similar fashion to undertaken here - after a prolonged period
of particle injection, they allowed the resultant synchrotron bubble 
to expand adiabatically. In their case, the light curve decays after only
$\sim$10~days, and they derive an expansion velocity of 400~km~s$^{-1}$.
In our case, the light curve decays on a much longer timescale which
naturally implies a much lower expansion velocity.
A velocity of only 12~km~s$^{-1}$ is surprisingly low for the outflow
of disk material at this distance from the Be star.
Waters et al. (1987)\nocite{wcl87} show that the velocity is likely
to be $\sim$100~km~s$^{-1}$ and indeed our optical observations of
this star indicate a velocity of $\sim$80~km~s$^{-1}$ at 20~$R_*$.
It is not clear, therefore, why our derived velocity is so low.

Synchrotron loss times, set by the magnetic field in the bubble,
are also relevant but on a longer timescale than the adiabatic losses.
As the synchrotron bubble evolves, the break frequency in
the emission spectrum moves to lower frequencies. Once the break
frequency becomes lower than the observing frequency, the
flux density at that frequency should decline rapidly.
The time, $t_b$, in days, at which the break frequency equals the observing
frequency can be approximated as
\beq
  t_b \approx 315 \,\, B^{-3/2} \nu^{-1/2}
  \label{eqn:tb}
\eeq
where $B$ is the magnetic field strength in Gauss and $\nu$ is in GHz.
Given that no further injection of electrons occurs after \p{+22}
and that the emission at 4.8 GHz appears to have decayed by
\p{+90} we can therefore estimate the decay of the emission at 8.4, 2.5, 1.4
and 0.84 GHz should occur 73, 118, 150 and 185 days after periastron
respectively. This is consistent with the lack of emission
at 8.4 GHz from \p{+63} onwards and the detection of unpulsed
emission at 1.4 and 0.8 GHz at least as late as \p{+113} and
\p{+132}. Using Equation~\ref{eqn:tb}, the magnetic field in the disc
where the synchrotron bubble is created is $\sim$1.6~G.
This is consistent with the result obtained by Ball et al. (1999).
Note that, as expected, this value of the magnetic field in the disc
significantly higher than the value we measured for the wind of
the Be star.

\section{Conclusions}
A series of observations of the 2000 periastron of the \psrbe{} system were
made at the ATCA at four frequencies. The pulsar binning mode
available with the ATCA correlator meant that
pulsed and unpulsed emission could be separated and we therefore
obtained light curves for the pulsar and unpulsed emission independently.
We were also able to obtain RMs and DMs for the pulsar.

The main features in the light curve from the 2000 periastron
are similar to the 1997 periastron, providing confirmation of
our earlier model. It does appear, however, that the pulsar
eclipse lasted longer and the unpulsed emission began and finished
later in 2000 than in 1997. This indicates that the disc has become
larger and/or denser in 3.5~yr.
As we obtained more data at late times than in the previous
periastron encounters, we found that adiabatic decay of the 
synchrotron bubble provided the best fit to the data.
The expansion rate of the bubble, $\sim$12~km~s$^{-1}$, is
significantly lower than expected but the magnetic field
value of 1.6~G consistent with measurements in other Be stars.

In addition, at \p{-46} the pulsar had the
highest measured astrophysical RM of $-14800 \pm 1800$ \rotm{}
implying a magnetic field of at least 6~mG in the Be star wind.

\section*{Acknowledgements}
The Australia Telescope is funded by the Commonwealth of 
Australia for operation as a National Facility managed by the CSIRO.
We thank A.~Karastergiou and B.~Koribalski for help with the observing
and A.~Melatos for useful discussions.

\bibliography{modrefs,psrrefs,crossrefs}
\bibliographystyle{mn}

\label{lastpage}
\end{document}